\documentclass[10pt,a4paper]{article}

\newcommand{\xa}{\mathbf{x}^a}
\newcommand{\xb}{\mathbf{x}^b}
\newcommand{\xt}{\mathbf{x}^t}
\newcommand{\xx}{\mathbf{x}}

\newcommand{\yy}{\mathbf{y}}
\newcommand{\eo}{\epsilon_o}
\newcommand{\eb}{\epsilon_b}
\newcommand{\ea}{\epsilon_a}

\newcommand{\oA}{\mathbf{A}}
\newcommand{\oB}{\mathbf{B}}
\newcommand{\oC}{\mathbf{C}}
\newcommand{\oH}{\mathbf{H}}
\newcommand{\oI}{\mathbf{I}}
\newcommand{\oK}{\mathbf{K}}
\newcommand{\oR}{\mathbf{R}}

\usepackage{graphicx}
\usepackage{dcolumn}
\usepackage{bm}

\begin{document}

\title{Unified nuclear core activity map reconstruction using heterogeneous
instruments with data assimilation}

\author{Bertrand Bouriquet $^1$ \footnote{bertrand.bouriquet@cerfacs.fr}
   \and Jean-Philippe Argaud $^{2}$
   \and Patrick Erhard $^2$
   \and Ang\'elique Pon{\c c}ot $^2$}

\maketitle

\footnotetext[1]{
Sciences de l'Univers au CERFACS, URA CERFACS/CNRS No~1875,
42 avenue Gaspard Coriolis,
F-31057 Toulouse Cedex 01 - France
}
\footnotetext[2]{
Electricit\'e de France,
1 avenue du G\'en\'eral de Gaulle,
F-92141 Clamart Cedex - France
}

\begin{abstract}

Evaluating the neutronic state of the whole nuclear core is a very important
topic that have strong implication for nuclear core management and for security
monitoring. The core state is evaluated using measurements. Usually, part of the
measurements are used, and only one kind of instruments are taken into account.
However, the core state evaluation should be more accurate when more
measurements are collected in the core.  But using information from
heterogeneous sources is at glance a difficult task.  This difficulty can be
overcome by Data Assimilation techniques. Such a method allows to combine in a
coherent framework the information coming from  model and the one coming from
various type of observations. Beyond the inner advantage to use heterogeneous
instruments, this leads to obtain a significant increasing of the quality of
neutronic global state reconstruction with respect to individual use of
measures. In order to present this approach, we will introduce here the basic
principles of data assimilation focusing on BLUE (Best Unbiased Linear
Estimation). Then we will present the configuration of the method within the
nuclear core problematic. Finally, we will present the results obtained on
nuclear measurement coming from various instruments.


\end{abstract}

\section{Introduction\label{sec:int}}

The knowledge of the neutronic state in the core is a fundamental point for the
design, the safety and the production processes of nuclear reactors. Due to the
crucial role of this information, considerable work has been conducted for long
time to accurately estimate the neutronic spatial fields. Spatial distribution
of power or activity in the whole core, or hottest point of the core, can be
derived from such spatial fields. These information allows mainly to check that
the nuclear reactor is working as expected in a very detailed manner, and that
it will remain in the operating limits when in production.

Two types of information can be used for the neutronic state evaluation.

First, the physical core specifications, including the nuclear fuel description,
make possible to build a numerical simulation of the system. Taking into account
neutronic, thermic and hydraulic spatial properties of the nuclear core, such
numeric models calculate in particular the useful reaction rates for the
physical analysis of the core state.

Secondly, various measurements can be obtained from in-core or out-of-core
detectors. Some detectors can measure neutron density, either locally or in
spatially integrated areas, other can measure temperature of the in-core water
at some points, etc. A lot of reliable measures comes from periodical flux maps
measured in each core reactor, at a periodicity of about one month. Then, all
these measurements do not have the same type of physical relation with the
neutronic activity, and also not the same accuracy. So it is uneasy to take into
account simultaneously all these heterogeneous measurements for the experimental
evaluation of the neutronic state in the core.

A lot of these measures are local, in determined fuel assemblies, and do not
give informations in un-instrumented areas of the core. The activity
distribution over the whole core is traditionally obtained through an
interpolation procedure, using the calculated field as first guess (a proxy) of
the ``real'' activity field corresponding to the measurements. In other words,
the activity value in un-instrumented areas is calculated as the weighted
average of the activity measures, using the calculated activity field to
interpolate. The power is then obtained from activity through an observation
operator, which depends only on core nominal physical specifications for the
periodical flux map measurements. This interpolation procedure gives already
good results, but some drawbacks remains in using only activity measures in a
deterministic interpolation procedure.

Both physical core specifications and real measurements are subject to some
uncertainties. Moreover, numerical assumptions, required to use the models, add
some inaccuracy. All these uncertainties are not used explicitly in the
interpolation procedure, but often used to qualify the \textit{a posteriori}
activity field build by the procedure. Moreover, the interpolation cannot take
into account, for example, heterogeneous instruments, or observed discrepancy of
some instruments.

Attempts have been made to overcome these limitations, mainly in two directions.
First, studies attempt to combine activity measurements and calculations through
least-squares derived methods (for example in \cite{Ezure88}), leading to the
most probable activity (or power) field on the whole core. These methods allow
to take into account heterogeneous measures, but are difficult to develop
because of their extreme sensitivity to the weighting factor in the combination
of measures and calculations. Secondly, explicit control of the error, in order
to reduce its importance, have been tried through the development of adaptive
methods to adjust coefficients in the calculation or the interpolation
procedure.

Some of these difficulties can be solved by using data assimilation. This
mathematical and numerical framework allows combining, in an optimal and
consistent way, values obtained both from experimental measures and from
\textit{a priori} models, including information about their uncertainties.
Commonly used in earth sciences as meteorology or oceanography, data
assimilation has strong links with inverse problems or Bayesian estimation 
\cite{era40,Tarantola87}. It is specifically tailored to solve such estimation
problems through efficient yet powerful procedures such as Kalman filtering or
variational assimilation \cite{Bouttier99}. Already introduced in nuclear field
\cite{Massart07,Bouriquet2010,Bouriquet2011}, it can be used both for field
reconstruction or for parameter estimation in a unified formalism.

Data assimilation can treat information coming from any type of measure
instruments, taking into account the way the measure is related with the
objective field to be reconstructed, such as neutronic activity here. Data
assimilation can further adapt itself to instrument configuration changes, and
for example the removal or the failure of an instrument. Moreover, the method
takes natively into account informations on instrumental or model
uncertrainties, introducing them \textit{a priori} through the data assimilation
procedure, and obtaining \textit{a posteriori} the reduced uncertrainties on the
reconstruction solution.

In this paper, we introduce the data assimilation method and how it address
physical field reconstruction. Then we will make a detailed description of the
various components that are used in data assimilation, and of the various type
of instruments we can use to make in-core neutronic activity measurement. Then
we present results with various instrument situations in nuclear core, obtained
on a set of true nuclear cores. 

\section{Data assimilation\label{sec:da}}

We briefly introduce the useful data assimilation key points, to understand
their use as applied here. But data assimilation is a wider domain, and these
techniques are for example the keys of nowadays meteorological operational
forecast. This is through advanced data assimilation methods that long-term
forecasting of the weather has been drastically improved in the last 30 years.
Forecasting is based on all the available data, such as ground and satellite
measurements, as well as sophisticated numerical models. Some interesting
information on these approaches can be found in the following basic references:
\cite{Talagrand97,Kalnay03,Bouttier99}. 

The ultimate goal of data assimilation methods is to be able to provide a best
estimate of the inaccessible ``true'' value of the system state, denoted $\xt$,
with the $t$ index standing for ``true''. The basic idea of data assimilation is
to put together information coming from an {\emph a priori} state of the system
(usually called the ``background'' and denoted $\xb$), and information coming
from measurements (denoted as $\yy$).  The result of data assimilation is called
the analysed state $\xa$ (or the ``analysis''), and it is an estimation of the
true state $\xt$ we want to find. Details on the method can be found in
\cite{Bouttier99} or \cite{Tarantola87}.

Mathematical relations between all these states need to be defined. As the
mathematical spaces of the background and of the observations are not necessary
the same, a bridge between them has to be build. This bridge is called the
observation operator $H$, with its linearisation $\oH$, that transform values
from the space of the background state to the space of observations. The
reciprocal operator is known as the adjoint of $H$. In the linear case, the
adjoint operator is the transpose $\oH^T$ of $\oH$.

Two additional pieces of information are needed. The first one is the
relationships between observation errors in all the measured points. They are
described by the covariance matrix $\oR$ of observation errors $\eo$, defined by
$\eo=\yy-H(\xt)$. It is assumed that the errors are unbiased, so that
$E[\eo]=0$, where $E$ is the mathematical expectation. $\oR$ can be obtained
from the known errors on the unbiased measurements. The second one is similar
with the relationships between background errors. They are described by the
covariance matrix $\oB$ of background errors $\eb$, defined by
$\epsilon_b=\xb-\xt$. This represents the {\it a priori} error, assuming it to
be unbiased. There are many ways to obtain this {\it a priori} and background
error matrices. However, they are commonly build from the output of a model with
an evaluation of its accuracy, and/or the result of expert knowledge. 

It can be proved, within this framework, that the analysis $\xa$ is the
Best Linear Unbiased Estimator (BLUE), and is given by the following formula:
\begin{equation}\label{xa}
\xa = \xb + {\bf K} \big(\yy-\oH\xb\big)
\end{equation}
where $\oK$ is the gain matrix \cite{Bouttier99}:
\begin{equation} \label{Kmat}
\oK = \oB\oH^T (\oH\oB\oH^T + \oR )^{-1}
\end{equation}
Moreover, we can get the analysis error covariance matrix $\oA$, characterising
the analysis errors $\ea=\xa-\xt$. This matrix can be expressed from $\oK$ as:
\begin{equation}
\oA = ( \oI - \oK \oH ) \oB
\end{equation}
where $\oI$ is the identity matrix.

The detailed demonstrations of those formulas can be found in particular in the
reference \cite{Bouttier99}. We note that, in the case of Gaussian distribution
probabilities for the variables, solving equation \ref{xa} is equivalent to
minimising the following function $J(\xx)$, $\xa$ being the
optimal solution:
\begin{equation}\label{J}
\begin{array}{lcc}
J(\mathbf{x}) &=& (\xx-\xb)^T \oB^{-1} (\xx-\xb) \medskip \\
              &+& \big(\yy-\oH\xx\big)^T \oR^{-1} \big(\yy-\oH\xx\big) \\
\end{array}
\end{equation}

We can make some enlightening comments concerning this equation \ref{J}, and
more generally on the data assimilation methodology. If we do extreme
assumptions on model and measurements, we notice that these cases are covered by
minimising $J$. First, assuming that the model is completely wrong, then the
covariance matrix $\oB$ is $\infty$ (or equivalently $\oB^{-1}$ is $0$). The
minimum of $J$ is then given by $\xa=\oH^{-1}\yy$ (denoting by $\oH^{-1}$ the
inverse of $\oH$ in the least square sense). It corresponds directly to
information given only by measurements in order reconstruct the physical field.
Secondly, on the opposite side, the assumption that measurements are useless
implies that $\oR$ is $\infty$. The minimum of $J$ is then evident: $\xa=\xb$
and the best estimate of the physical field is then de calculated one. Thus,
such an approach covers whole range of assumptions we can have with respect to
models and measurements. 

\section{Data assimilation method parameters}

The framework of the study is the standard configuration of a 1300 electrical~MW
Pressurized Water Reactor (PWR1300) nuclear core. Our goal is to reconstruct the
neutronic fields, such as the activity, in the active part of the nuclear core.
For that purpose, we use data assimilation. To implement such methodology, we
need both simulation codes and measures. For the simulation code, we use
standard EDF calculation code for nuclear core simulation, in a typical
configuration. The results are build on a set of 20 experimental neutronic flux
maps measured on various PWR1300 nuclear cores. Such measurements are done
periodically (about each month) on each nuclear core. These different
measurement situations are chosen for their representativeness, in order that
statistical results covers a wide range of situations and can have some sort of
predictability property.

\subsection{The backgroud and the measurements}

A standard PWR1300 nuclear core has $193$ fuel assemblies within. For the
calculation, those assemblies are each considered as homogeneous, and are
divided in $38$ vertical levels. Thus, the state field $\xx$ can be represented
as a vector of size $193 \times 38=7334$.

\begin{figure}[!ht]
\begin{center}
  \includegraphics[width=\textwidth]{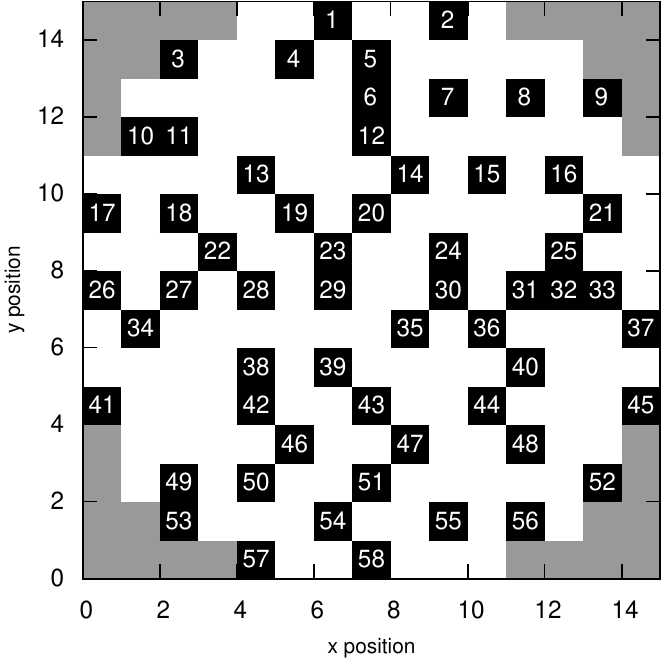}

  \caption{The Mobile Fission Chambers (MFC) instruments within the nuclear core
  are localised in assemblies in black, in a horizontal slice of the core. The
  assemblies without instrument are marked in white, and the reflector is in
  grey. \label{fig:figCFM}}

\end{center}
\end{figure}

\begin{figure}[!ht]
\begin{center}
  \includegraphics[width=\textwidth]{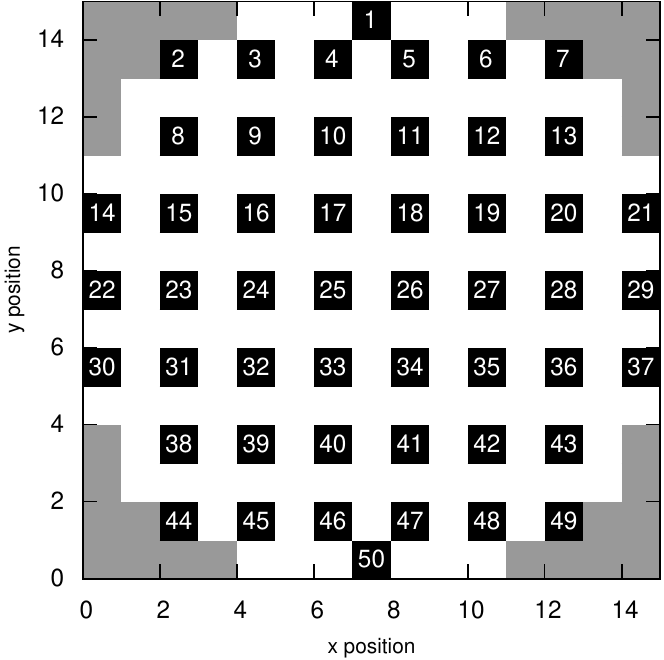}

  \caption{The Thermocouples (TC) instruments within the nuclear core are
  localised above assemblies in black, in a horizontal slice of the core. The
  assemblies without instrument are marked in white, and the reflector is in
  grey. \label{fig:figTC}}

\end{center}
\end{figure}

The measurements come from instruments that can be located on horizontal 2D maps
of the core. There are three types of instrument that are usually used to
monitor the nuclear power core:
\begin{itemize}
\item Mobile Fission Chambers (MFC), which are inside the active nuclear core,
      for which the locations are presented on the figure \ref{fig:figCFM},
\item Thermocouples (TC), which are above the active nuclear core, for which the
      locations are presented on figure \ref{fig:figTC},
\item fixed ex-core detector locations.
\end{itemize}

The data coming from the ex-core detectors are continuous in time and are very
efficient for security purpose, which is their main goal. Their purpose is to
continuously monitor the core, but not to measure accurately the neutronic
activity at each fine flux map. So, their measures are too crude for being
interesting on a fine reconstruction of the inner core activity map. Thus, we do
not take into account information coming from those data.

\begin{figure}[!ht]
\begin{center}
  \includegraphics[width=\textwidth]{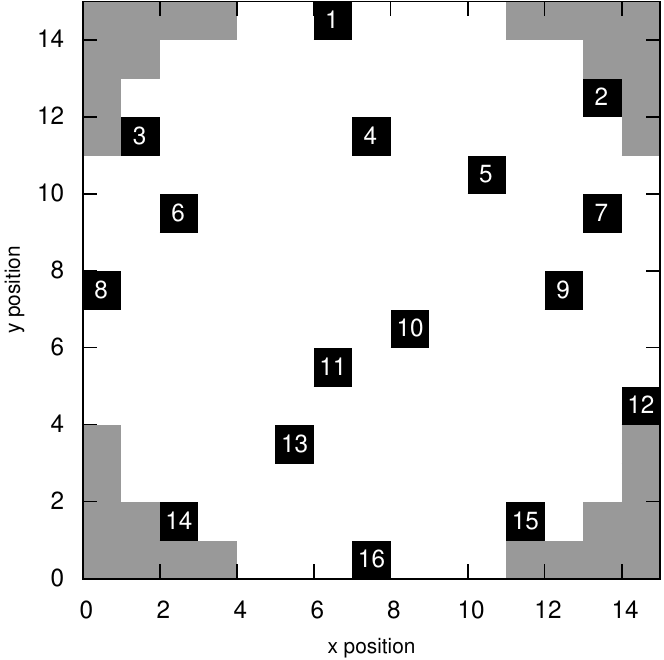}

  \caption{The idealized Low Granularity MFC (LMFC) instruments within the
  nuclear core are localised in assemblies in black, in a horizontal slice
  of the core. The assemblies without instrument are marked in white, and the
  reflector is in grey. \label{fig:figCol}}

\end{center}
\end{figure}

All these types of instrumentation (MFC, TC, ex-core) can be found on any power
plants. For the purpose of this study, we add artificially an extra type of
detector, described as idealized Low Granularity MFC (named here LMFC). The
measurement attributed to the LMFC are build artificially from the information
given by the MFC. Thus they are replacing the MFC on the given LMFC locations.
The evaluation of LMFC response is calculated from the MFC measured neutron
flux, assuming a different physic process, and a lower granularity. The lower
granularity assumption done on the LMFC induces a partial integration of the
results of the MFC over a given area. Of course, the physical process involved
to make a measurement being different, the resolution of LMFC will be different
from the one from MFC. We take $16$ of those instruments. They are located in
various area of the core, replacing MFC, to try to make a representative array
of measurement as in is shown on figure \ref{fig:figCol}.

The main characteristics of the instruments, as there number in the core, the
number of considered vertical levels, and the size of the part of the
observation vector $\yy$ associated with the particular instrument type, are
reported in table \ref{tab:intrucarac}. The size of the final observation vector
is given by summing the size of all the individual $\mathbf{y}$ vector of the
instruments used.

\begin{table}[!ht]
  \centering
  \begin{tabular}{|c|c|c|c|}
    \hline
    \textit{Instrument}  & \textit{Locations} & \textit{Vertical} & \textit{Size of the $\yy$} \\
       \textit{type}     &  \textit{number}   &  \textit{levels}  &  \textit{vector part}      \\ \hline
       MFC      &   $42$    &   $38$   & $1596$            \\ \hline
       LMFC     &   $16$    &    $8$   &  $128$            \\ \hline
       TC       &   $50$    &    $1$   &   $50$            \\ \hline
  \end{tabular}
  \caption{Main characteristics of the instruments used for data assimilation.
  These characteristics remains the same, either in mono-instrumented cases or
  in multi-instrumented ones. \label{tab:intrucarac}}
\end{table}
\subsection{The observation operator $\oH$}

The $H$ observation operator is the composition of a selection and of a
normalisation procedure, and can be build independently for each instrument.
So there is one individual $\oH$ matrix observation operator by instrument type.
The complete $\oH$ matrix observation operator is the concatenation, as a
bloc-diagonal matrix, of all the individual matrix for each instrument.

Each observation operator is basically a selection matrix, that choose in the
model space a cell that is involved in a measurement in the observation space.
In addition, a weight, according to the size of the cell, is affected to the
selection. As experimental data are normalised, this selection matrix is
multiplied by a normalisation matrix that represents the effect of the cross
normalisation of the data. This observation matrix is a $(7334 \times \sum d_i)$
matrix, where $d_i$  is the size of the part of the observation vector $\yy$ for
each instrument involved in assimilation, as reported in table
\ref{tab:intrucarac}.

\subsection{The background error covariance matrix $\oB$}

The $\oB$ matrix represents the covariance between the spatialised errors for
the background. The $\oB$ matrix is estimated as the double-product of a
correlation matrix $\oC$ by a diagonal scaling matrix containing standard
deviation, to set variances.  

The correlation $\oC$ matrix is built using a positive function that defines the
correlations between instruments with respect to a pseudo-distance in model
space. Positive functions allow (via Bochner theorem) to build symmetric defined
positive matrix when they are used as matrix generator (for theoretical insight,
see reference documents \cite{Matheron70} and \cite{Marcotte08}). Second Order
Auto-Regressive (SOAR) function is used here. In such a function, the amount of
correlation depends from the euclidean distance between spatial points in the
core. The radial and vertical correlation lengths (denoted $L_r$ and $L_z$
respectively, associated to the radial $r$ coordinate and the vertical $z$
coordinate) have different values, which means we are dealing with a global
pseudo euclidean distance. The used function can be expressed as follow:
\begin{equation}  
C(r,z) = \left(1+\frac{r}{L_r}\right) \left(1+\frac{|z|}{L_z}\right)
         \exp{\left(-\frac{r}{L_r}-\frac{|z|}{L_z}\right)}.
\label{eqB}
\end{equation}
The matrix $\oC$ obtained from the above Equation \ref{eqB} is a correlation
one. It can be multiplied (on left and right) by a suitable diagonal standard
deviation matrix, to get covariance matrix. If the error variance is spatially
constant, there is only one coefficient to multiply $\oC$. This coefficient is
obtained here by a statistical study of difference between the model and the
measurements in real case. In real cases, this value is set around a few
percent.

The size of the background error covariance matrix $\oB$ is related to the size
of model space, so it is $(7334 \times 7334)$ here.

\subsection{The observation error covariance matrix $\oR$} 

The observation error covariance matrix $\oR$ is approximated by a simple
diagonal matrix. It means we assume that no significant correlation exists
between the measurement errors of all the instruments. A usual modelling
consists in taking the diagonal values as a percentage of the observation
values. This can be expressed as:
\begin{equation}  
\oR_{jj} = \left( \alpha \yy_j \right)^2, \quad \forall j.
\label{eqR}
\end{equation}
The $\alpha$ parameter is fixed according to the accuracy of the measurement and
the representative error associated to the instrument. It is the same for all
the diagonal coefficients related with one instrument. Its values is only
depending of the type of instrument we are dealing with. The $\alpha$ value can
be determined by both statistical method and expert opinion about the
measurement quality. In the present paper, we will use arbitrary value for the
$\alpha$.

The size of the $\oR$  matrix is related to the size of the observation space,
so it is $(\sum_i d_i \times \sum_i d_i)$ where $d_i$ is the size of the
observation vector of each instrument $i$ involved in assimilation, as reported
in the table \ref{tab:intrucarac}.

\section{Results on data assimilation using only one type of instrument
\label{sec:nono}}

The first results are showing the quality of the reconstruction as a function of
the various types of instruments that are taken into account for reconstruct the
activity of the core. 

The experimental data are a set of measurement on the 38 levels of the all the
instrument locations inside of the core. Thus, to evaluate the quality of the
reconstruction of the physical fields with one type of instrument, we look for
the misfit $(\yy-\oH\xa)$ at measurement locations (by other instruments) that
are not involved in the assimilation process. The number of locations, where
there is a measurement and that is not involved in data assimilation procedure,
so where the misfit $(\yy-\oH\xa)$ is calculated, are synthesised in the table
\ref{tab:accuracy}, as well as the accuracy associated to each instrument
through the $\alpha$ parameter of equation \ref{eqR}.

\begin{table}[!ht]
  \centering
  \begin{tabular}{|c|c|c|}
    \hline
    \textit{Instrument} & \textit{Number of misfit} & $\alpha$ \textit{value} \\
       \textit{type}    & \textit{calculations locations} & ~                       \\ \hline
       MFC              & $16$               &  $1\%$                  \\ \hline 
       LMFC             & $42$               &  $2\%$                  \\ \hline 
       TC               & $58$               &  $3\%$                  \\ \hline 
  \end{tabular}
  \caption{Number of misfit calculation locations used for each instrument type
  considered individually for data assimilation, and arbitrary accuracy assumed
  in the present studies for each type of instrument. \label{tab:accuracy}}
\end{table}

Thermocouples being a fully integral measurement outside of the active core, we
can do the misfit $(\yy-\oH\xa)$ evaluation of the reconstruction in all the
locations of MFC/LMFC. 

For each data assimilation procedure associated with an instrument, we calculate
the Root Mean Square (RMS, which is the norm) of the misfit $(\yy-\oH\xa)$ on
all the misfit calculation locations. To synthesize the value for one set of
measurement, we take the mean value of the misfit $(\yy-\oH\xa)$ on each of the
$38$ levels of the core, which leads to a horizontally averaged value of the
misfit $(\yy-\oH\xa)$. In sake of more general behaviour, we take $20$ set of
flux map measurements, with various settings and ageing of PWR1300 nuclear
cores. Then we proceed to the calculation of the mean value on all those set of
measurement. 

Globally all the results present strong misfit $(\yy-\oH\xa)$ on the upper and
lower levels, which is a known effect mainly due to the axial reflector
modelling. In addition, the centre part of the reactive core in the nuclear
plant is also the region where the neutron flux is the most intense, so the hot
spot of the activity field in the core is expected to be in this region. Thus,
the next plots are restricted to the centre of the core.

\begin{figure}[!ht]
\begin{center}
  \includegraphics[angle=-90,width=\textwidth]{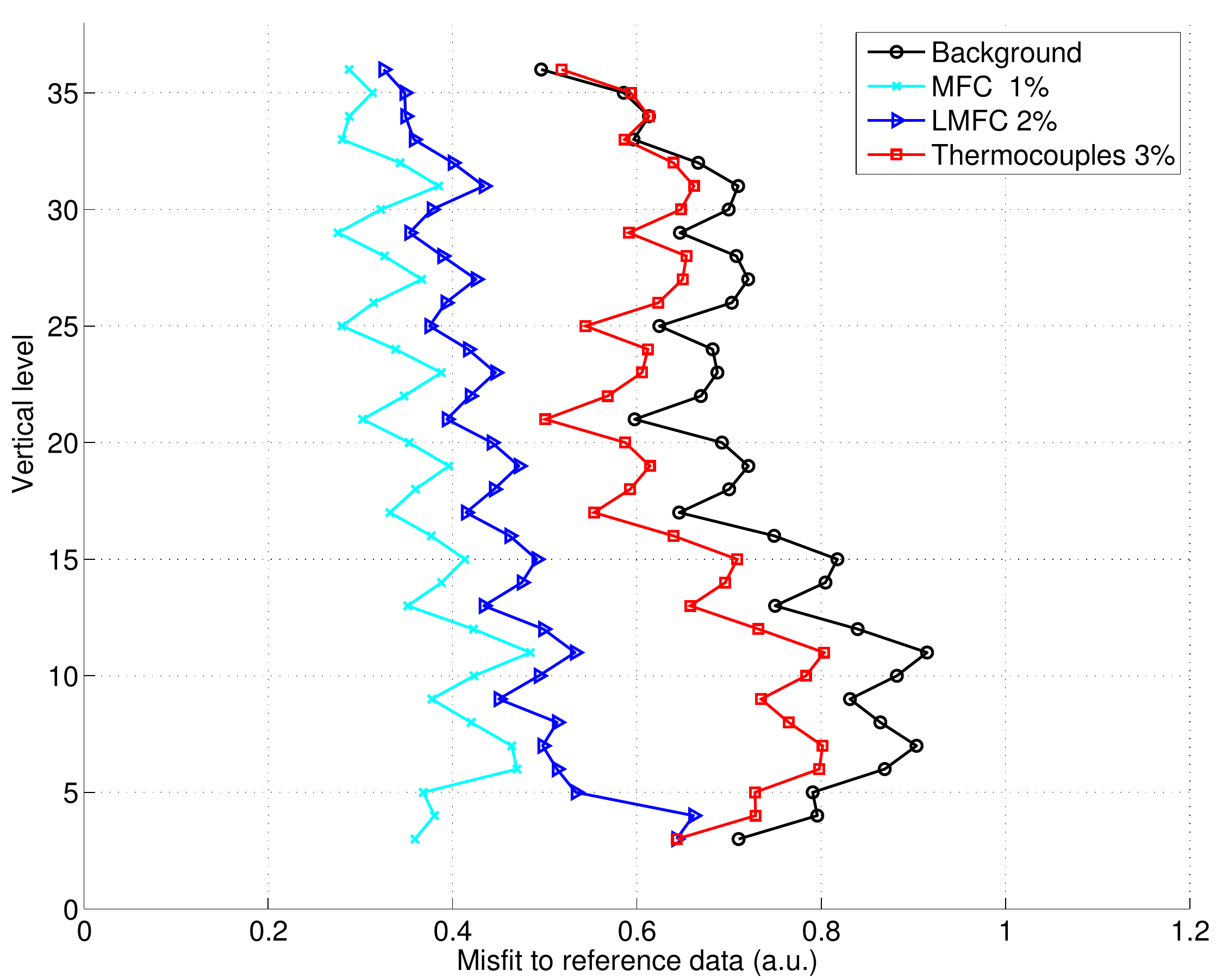} 
  \caption{Vertical misfit for various kind of intruments, measured by the RMS
  of the differences between the analysis and the measurements at unused
  locations (see text for details). \label{fig:fig1}}
\end{center}
\end{figure} 

Figure \ref{fig:fig1} shows the axial misfit measured by the standard RMS of the
difference between analysis and measurements, in arbitrary units, on all the
data assimilation unused locations for the various types of instrument studied. 
The oscillating behaviours, that are barely noticeable on all the curves, come
from the different material that are within a core level. Some levels are
containing mechanical structure of the core, thus these are more neutron
absorptive. 

We noticed, as expected, that the reconstruction coming from the thermocouples
(TC) is the closest to the background, due to their integral measurement
property and their lower accuracy. Moreover, an improvement of the accuracy does
not improve dramatically the quality of the core state evaluation, mainly due to
the integral measurement property.  

The Low Granularity MFC (LMFC) are showing a good reconstruction of the physical
field in the whole core, despite the not so good accuracy and the limited number
of measurement. The increase of the misfit from $(\yy-\oH\xb)$ to $(\yy-\oH\xa)$
for the lower part of the core is easily explainable by the chosen locations of
LMFC that are in core. This part, near the border of the core, does not get
enough measurements to be very accurately reconstructed.  

The reconstruction using only MFC is, also as expected from accuracy values, the
best one. The data assimilation procedure leads to half the misfit
$(\yy-\oH\xb)$ observed when only using the model.

From results coming from MFC and LMFC, we notice that, within the hypothesis and
the chosen modelling of the integration, TC measurement are permitting only a
crude evaluation of the core state. 

\section{Results on data assimilation with heterogeneous instruments}

This section describes results using different instruments together in the data
assimilation procedure.

\begin{figure}[!ht] 
\begin{center} 
  \includegraphics[angle=-90,width=\textwidth]{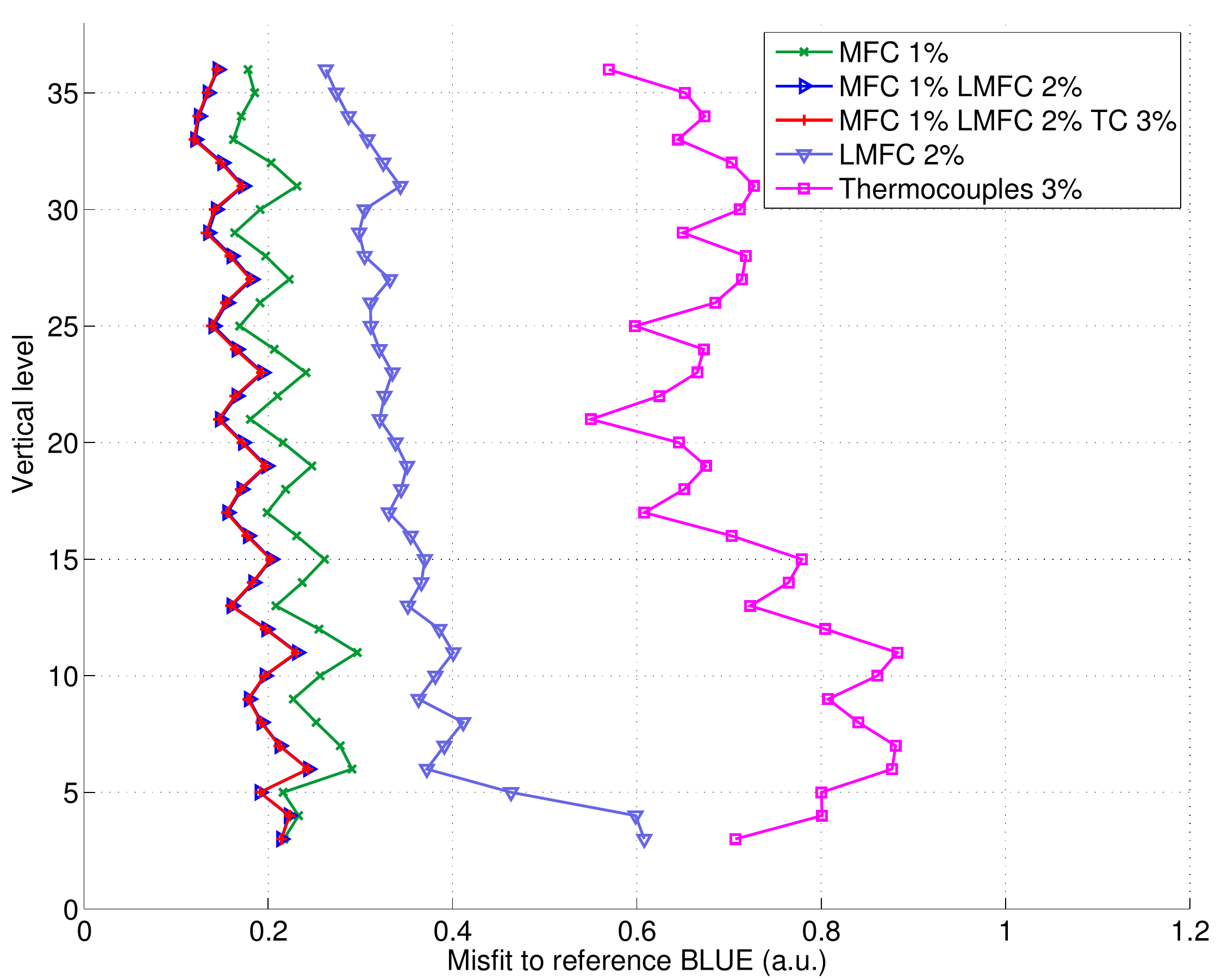} 
  \caption{Axial misfit for various kind of instrumental configurations,
  measured by the RMS of the difference between the analysis and the reference
  calculation. \label{fig:fig2}}
\end{center}
\end{figure}

In this case, we cannot take as reference the measures points taken apart, as
when we are studding individual instrument as presented in section
\ref{sec:nono}. Thus, we choose to make an analysis with data assimilation using
all the available measures in the core, on the $58$ locations, and using a $1\%$
accuracy. Then we evaluate the misfit $(\yy-\oH\xa)$ with respect to this
reference calculation in all the assemblies where no MFC are present, which
means $135$ locations ($193$ total assembly minus $58$ instrumented locations).
The calculated values at those instrumented locations allow to benchmark the
quality of the reconstruction. On this misfit $(\yy-\oH\xa)$, we calculate the
RMS per horizontal level as previously, and take the average over some $20$
selected flux map measurements. The results are presented on figure
\ref{fig:fig2}.

The figure \ref{fig:fig2} presents RMS per horizontal level of the misfit
calculated for various instrument taken alone or in conjunction. We notice that
the results have the same behaviour as the one we got in figure \ref{fig:fig1}.
This means that the reference we chose to evaluate the quality of data
assimilation using several instrument types can be considered as reliable.

Looking at the successive addition of instrument, we notice that addition of the
LMFC have a important effect, the MFC+LMFC configuration presenting an
improvement with respect to the configuration with only MFC. However, adding
thermocouple to this configuration is not really helpful, and the improvement is
not really noticeable on this figure \ref{fig:fig2}. 

These results highlight a number of important points on data assimilation
methodology. On one hand, when only very few measurements are available, they
are very helpful and allow a fairly good improvement of reconstructed physical
field. On the other hand, when a lot of measurements are available, adding a few
more, or with a lower accuracy as thermocouples, do not change dramatically the
result of field reconstruction by data assimilation. On overall, this shows that
data assimilation technique is doing the best use of experimental information
provided to the procedure. Those results are comforting the ones found on the
robustness of the evaluation of the nuclear core by data assimilation, when only
MFC are used as presented in \cite{Bouriquet2011}. Moreover, as expected in
data assimilation technique, the use of heterogeneous instrument is transparent
within the method. 

\section{Conclusion}

The use of data assimilation has already been proved to be efficient to
reconstruct fields in several domains, and recently in neutronic activity field
reconstruction for nuclear core. The present paper demonstrates that, within the
data assimilation framework, information coming from heterogeneous sources can
be used in a transparent way without making any adjustment to the method.

Looking at the various types of instruments, we have (MFC, LMFC and TC) we
notice that the influence they have on reconstruction depends on three
parameters:
\begin{itemize}
\item the granularity of each type of instrument, that is the density of
instruments, their integral measurement property and their repartition all over
the core,
\item the accuracy of each instrument, possibly with respect to the accuracy of
the others,
\item and the global instrumentation configuration, that is the complex
repartition of all instruments.
\end{itemize}
Those conclusion arise from comparison of the various instruments, using them
individually or together, in various instrumental configurations.

Data assimilation gives a very efficient and adaptable framework in order to
take the best from both experimental data and model. Moreover, this can be done
without heterogeneities constrains between instruments. This technique is used
here in a very elementary way, but it opens the door to many developments, for
example in systematic data analysis, models comparison, in dynamic modelling,
etc. for nuclear reactors. 


\bibliography{bibliographie}

\begin{thebibliography}{10}
\expandafter\ifx\csname url\endcsname\relax
  \def\url#1{\texttt{#1}}\fi
\expandafter\ifx\csname urlprefix\endcsname\relax\def\urlprefix{URL }\fi
\expandafter\ifx\csname href\endcsname\relax
  \def\href#1#2{#2} \def\path#1{#1}\fi

\bibitem{Ezure88}
H.~Ezure, Estimation of most probable power distribution in bwrs by least
  squares method using in-core measurements, Journal of Nuclear Science and
  Technology 25~(9) (1988) 731--740.

\bibitem{era40}
S.~M. Uppala, {\it et al.}, The \uppercase{ERA}-40 re-analysis, Quaterly
  Journal of the Royal Meteorological Society 131~(612, Part B) (2005)
  2961--3012.

\bibitem{Tarantola87}
A.~Tarantola, Inverse Problem: Theory Methods for Data Fitting and Parameter
  Estimation, Elsevier, 1987.

\bibitem{Bouttier99}
F.~Bouttier, P.~Courtier, Data assimilation concepts and methods,
  Meteorological training course lecture series, ECMWF (March 1999).

\bibitem{Massart07}
S.~Massart, S.~Buis, P.~Erhard, G.~Gacon, Use of \uppercase{3DVAR} and
  \uppercase{K}alman filter approaches for neutronic state and parameter
  estimation in nuclear reactors, Nuclear Science and Engineering 155~(3)
  (2007) 409--424.

\bibitem{Bouriquet2010}
B.~Bouriquet, J.-P. Argaud, P.~Erhard, S.~Massart, A.~Pon\c{c}ot, S.~Ricci,
  O.~Thual, Differential influence of instruments in nuclear core activity
  evaluation by data assimilation, Nuclear Instruments and Methods in Physics
  Research Section A 626-627 (2011) 97--104.

\bibitem{Bouriquet2011}
B.~Bouriquet, J.-P. Argaud, P.~Erhard, S.~Massart, A.~Pon\c{c}ot, S.~Ricci,
  O.~Thual, Robustness of nuclear core activity reconstruction by data
  assimilation, Nuclear Instruments and Methods in Physics Research Section A
  629~(1) (2011) 282--287.

\bibitem{Talagrand97}
O.~Talagrand, Assimilation of observations, an introduction, Journal of the
  Meteorological Society of Japan 75~(1B) (1997) 191--209.

\bibitem{Kalnay03}
E.~Kalnay, Atmospheric Modeling, Data Assimilation and Predictability,
  Cambridge University Press, 2003.

\bibitem{Matheron70}
G.~Matheron, La th\'eorie des variables r\'egionalis\'ees et ses applications,
  Cahiers du Centre de Morphologie Math\'ematique de l'ENSMP, Fontainebleau,
  Fascicule 5, 1970.

\bibitem{Marcotte08}
D.~Marcotte, G\'eologie et g\'eostatistique mini\`eres (lecture notes) (2008).

\end{thebibliography}
\bibliographystyle{elsarticle-num}

\end{document}